\documentclass[11pt,a4paper]{article}
\pdfoutput=1

\usepackage{jheppub}

\usepackage[T1]{fontenc}
\usepackage[utf8]{inputenc}
\usepackage{amsmath}
\usepackage{amssymb}
\usepackage{hyperref}
\usepackage{graphicx} 
\usepackage{mciteplus}

\title{Geometrical CP violation with a complete fermion sector}
\author[a]{Ivo de Medeiros Varzielas}
\emailAdd{ivo.de@unibas.ch}
\affiliation[a]{\small Department of Physics, University of Basel,\\ Klingelbergstr. 82, CH-4056 Basel, Switzerland}
\author[b]{Daniel Pidt}
\emailAdd{daniel.pidt@tu-dortmund.de}
\affiliation[a,b]{\small Facult\"at f\"ur Physik, Technische 
      Universit\"at Dortmund\\ D-44221 Dortmund, Germany}
      
\abstract{
We consider $\Delta(27)$ models featuring geometrical CP violation, and analyse several structures that can obtain the observed lepton masses and mixing.
The leptonic structures considered are entirely consistent with structures that reproduce the experimental data in the quark sector.
This constitutes an existence proof of geometrical CP violation models accounting for the masses and mixing of all fermions.
}

\preprint{DO-TH 13/20}

\begin{document}
\maketitle

\section{Introduction}

Geometrical CP violation (GCPV) was initially proposed in \cite{Branco:1983tn},
defined as a situation where CP is spontaneously violated by a vacuum expectation value (VEV) and the complex phase of that VEV takes a specific value that is independent of the magnitudes of the various parameters of the scalar potential (these specific complex phases are then referred to as ``calculable phases'').
The concept has been developed further in \cite{deMedeirosVarzielas:2011zw, Varzielas:2012nn, Varzielas:2012pd, Bhattacharyya:2012pi, Ivanov:2013nla, Varzielas:2013zbp, Ma:2013xqa, Varzielas:2013sla}.
Additionally, CP violation in the lepton sector has also been considered independently of the scalar potential for example in the two very general analyses \cite{Feruglio:2012cw} and \cite{Holthausen:2012dk} (the latter also briefly addresses the $\Delta(27)$ case of GCPV and calculable phases as defined in \cite{Branco:1983tn}).
The most explored GCPV scenario is an extension of the Standard Model (SM) extended with a CP symmetry and with $\Delta(27)$. For a $\Delta(27)$ triplet $\phi$, as long as cubic terms are not allowed, the renormalisable potential has the VEV \cite{Branco:1983tn}
\begin{equation}
\langle \phi \rangle = v(\omega,1,1) \,,
\label{eq:VEV}
\end{equation}
with $\omega \equiv e^{i 2 \pi/3}$.
Not all complex VEVs violate CP, as in many cases they can be eliminated by a rephasing symmetry of the
potential (see e.g. \cite{Varzielas:2012pd, Ivanov:2013nla}),
but one can verify explictly that the corresponding transformation $U$ that relates eq.(\ref{eq:VEV}) to its conjugate is not a symmetry of the potential:
\begin{equation}
\langle \phi \rangle \longrightarrow \langle \phi \rangle^{\ast} = U_{ij} \langle H_{j} \rangle \,,
\label{eq:U}
\end{equation}
which is sufficient to prove that it spontaneously violates CP regardless \cite{Branco:1983tn}.
The absence of cubic terms has been traditionally guaranteed because the $\Delta(27)$ triplet is charged as a doublet under the SM (see discussion in \cite{Varzielas:2012pd}).
It is very constraining to use only this $\Delta(27)$ triplet VEV to produce viable patterns of fermion masses and mixing, although some promising leading order structures can be obtained \cite{deMedeirosVarzielas:2011zw}. A viable GCPV framework focusing on the quark sector was presented first in \cite{Bhattacharyya:2012pi} and improved recently in \cite{Varzielas:2013sla}.

The lepton sector has also been considered, with some promising structures suggested by \cite{Ma:2013xqa}.
Here we turn our focus to the lepton sector, investigating the viability of several possibilities.
One of our aims is to develop a leptonic sector that is not only viable by itself, but fully compatible with the quark sector structures proposed in \cite{Bhattacharyya:2012pi, Varzielas:2013sla}. In doing so we intend to keep the $U(1)_F$ Froggatt-Nielsen (FN) symmetry \cite{Froggatt:1978nt} introduced in \cite{Varzielas:2013sla}.
By finding viable leptonic structures that are consistent with viable quark structures, we provide an existence proof of a complete model of fermion masses and mixing featuring GCPV.

\section{General considerations}

One obstacle that the leptonic sector presents is that the simplest extensions of the $\Delta(27)$ GCPV framework do not readily work.
In order to understand this, we briefly review the relevant properties of $\Delta(27)$: it is a discrete non-Abelian subgroup of $SU(3)$ with 27 elements.
Its irreducible representations are
one-dimensional  $1_{ij}$ ($i,j=0,1,2$ for a total of nine) and the two
three-dimensional $3_{01}$ and $3_{02}$ which act like anti-triplet and triplet.
The contractions are 
\begin{equation}
3_{0i} \times 3_{0i} = 3_{0j} + 3_{0j} + 3_{0j}\,,
\end{equation}
with the products between two antitriplets $A$, $B$ being three triplets (and vice-versa) built as
$(A_1 B_1, A_2 B_2, A_3 B_3)$, $(A_2 B_3,A_3 B_1,A_1 B_2)$, and $(A_3 B_2,A_1 B_3,A_2 B_1)$,
and
\begin{equation}
3_{01} \times 3_{02} = \sum_{i,j} 1_{ij}\,,
\end{equation}
with the product between an anti-triplet and triplet resulting in the singlets,
the invariant $1_{00}$ being constructed from a $3_{01}$ $A$ and a $3_{02}$ $B$ as $A_1 B_1 + c.p.$, with
$A_1 B_2 + c.p.$ transforming as $1_{01}$ and 
$A_1 B_3 + c.p.$ as $1_{02}$  ($c.p.$ denotes cyclic permutations of the indices).

If $H$ and $L$ are respectively the scalar and lepton SM doublets, we wish to construct
charged lepton invariants of type $H L \tau^c$ and
effective neutrino mass terms are $H^\dagger H^\dagger L L$. \footnote{
We refer to non-renormalisable terms without the necessary mass suppressions, which remain implicit. They are to be understood as the mass of some kind of messenger fields associated with some type of seesaw mechanism or with the ultraviolet completion of the flavour model, as in e.g. \cite{Varzielas:2010mp, Varzielas:2012ai, Luhn:2013lkn, Antusch:2013rla, Ding:2013bpa}.}
If $H$ transforms as a triplet under $\Delta(27)$ (we assume a $3_{01}$, without loss of generality), then in order to have invariant neutrino mass terms $H^\dagger H^\dagger L L$, $L$ must transform in the same triplet representation of $\Delta(27)$ (e.g. both are $3_{01}$). This then requires the $SU(2)$ singlet charged leptons to also transform as the same representation of $\Delta(27)$, which leads to mass structures that are not viable \cite{Branco:1983tn, deMedeirosVarzielas:2011zw}.
By extending the field content we can avoid this obstacle, in fact \cite{Ma:2013xqa} obtained viable lepton masses simply by introducing additional $SU(2)$ doublets $\zeta$ transforming as $\Delta(27)$ singlets: then $\zeta H^\dagger L L$ gives mass to the neutrinos, where $L$ should now be the conjugate triplet (a $3_{02}$), with $e^c$, $\mu^c$ and $\tau^c$ transforming as singlets.
Here we explore a different framework where we instead assign $H$ as a trivial $\Delta(27)$ singlet and add $\phi$, a $3_{01}$ and SM singlet, as shown in Table \ref{ta:content}. This separates the scalars responsible for breaking the SM, $H$, and $\phi$, that breaks both $\Delta(27)$ and CP. While this has the disadvantage that the interesting scalar sector properties discussed in \cite{Bhattacharyya:2012pi} no longer apply to the $SU(2)$ doublet scalars, conversely it allows the scale of CP breaking to be higher, which may be necessary for the generation of the baryon asymmetry of the universe (e.g. through leptogenesis).

\begin{table}
  \centering
  \begin{tabular}{|c||ccccc|cccc|cccc|}
    \hline
    & $Q_1$ & $Q_2$ & $Q_3$ & $u^c$ & $d^c$ & $L$ & $e^{c}$ & $\mu^{c}$ & $\tau^{c}$ & $H$ & $\phi$ & $\varphi$ & $\theta$ \\
    \hline\hline
    $\Delta(27)$  & $1_{00}$ & $1_{00}$ & $1_{02}$ & $3_{02}$ & $3_{02}$ &  $3_{01}$ &  $1_{00}$ &  $1_{00}$ & $1_{02}$ & $1_{00}$ & $3_{01}$ & $1_{00}$ & $1_{02}$ \\
    $U(1)_F$ & $3$ & $2$ & $0$ & $-p$ & $-p$ & $f_{L}\left( p \right)$ & $f_{e}\left( p \right)$ & $f_{\mu}\left( p \right)$ & $f_{\tau}\left( p \right)$ & $0$ & $p$ & $-1$ & $-2$ \\
    \hline
  \end{tabular}
  \caption{The symmetry and field content of the models. \label{ta:content}}
\end{table}

The separation of $H$ and $\phi$ also allows us to place $L$ as a $3_{01}$ and build $H L \phi^\dagger \tau^c $ consistently with neutrino mass terms $H^\dagger H^\dagger (L L \phi)$ or $H^\dagger H^\dagger (L L \phi^\dagger \phi^\dagger)$. Before proceeding into a detailed analysis of possible lepton structures, two issues must be addressed. As mentioned already, eq.(\ref{eq:VEV}) is valid in the absence of cubic $\phi$ terms in the potential. As $\phi$ is not an $SU(2)$ doublet, this has to be enforced by a different symmetry. As we are using a FN symmetry, we require that $\phi$ is not neutral under it and that the specific assignment does not allow any renormalisable terms cubic in $\phi$.
The second issue is that we need to make a small adaptation compared to the quark structures of \cite{Bhattacharyya:2012pi, Varzielas:2013sla}. Whereas there the invariants are $H^\dagger Q u^c$ and $H Q d^c$, with $H$, $u^c$ and $d^c$ respectively $3_{01}$, $3_{01}$ and $3_{02}$ under $\Delta(27)$, they should now be 
$H^\dagger Q u^c \phi$ and $H Q d^c \phi$, with $H$, $u^c$, $d^c$ and $\phi$ respectively $1_{00}$, $3_{02}$, $3_{02}$ and $3_{01}$ under $\Delta(27)$. Note that although this does not make a difference for the masses and CKM mixing, it does change the up mass matrix: $u^c$ is now the $3_{02}$ in the invariant contraction (with $\phi$, not $\phi^\dagger$). So the leading order
\begin{equation}
M_u = v \begin{pmatrix}
	x_{1} \omega & x_{1} & x_{1} \\
	x_{2} \omega & x_{2} & x_{2} \\
	x_{3} & x_{3} & x_{3} \omega 
\end{pmatrix} \,,
\end{equation}
can be compared with \cite{Varzielas:2013sla} where the equivalent structure has $\omega^2$ instead of $\omega$, and the $(32)-(33)$ entries are swapped.
Similarly, the invariants of type $H Q d^c (H H^\dagger)$ that were required to generate the complex phase in the CKM matrix are easily replaced by $H Q d^c \phi (\phi \phi^\dagger)$ and the quark sector remains equally viable. For further details about the specific structures we refer the reader to \cite{Bhattacharyya:2012pi, Varzielas:2013sla}.

With a continuous $U(1)_F$ (or discrete $Z_N$) FN symmetry, the mass hierarchies of the fermions can be addressed and the problematic scalar couplings of the GCPV scalar $\phi$ with non-trivial $\Delta(27)$ singlets can be forbidden. We will revisit the extended scalar sector in section \ref{sec:scalar}. The FN fields $\varphi$ and $\theta$ are SM gauge singlets, and charged under $\Delta(27)$ and $U(1)_F$ as $1_{00}$, $-1$ and $1_{02}$, $-2$. The Higgs field $H$ that breaks the SM $SU(2)$ does not transform under $\Delta(27)$, instead the $\Delta(27)$ triplet scalar, $\phi$ has some charge under FN, which we denote as $p$ in Table \ref{ta:content} and only specify it when we discuss model implementations. \footnote{
For GCPV, the requirement is that $\phi^3$, $\phi^3 \theta$, $\phi^3 \theta^\dagger$, $\phi^3 \varphi$ and $\phi^3 \varphi^\dagger$ are not invariant.}

\section{Leptonic structures \label{sec:leptons}}

With viable quark sector structures analogous to those of \cite{Bhattacharyya:2012pi, Varzielas:2013sla} guaranteed, we focus now on the different possibilities for the leptonic sector.
In general we choose to have $L$ transform as a triplet.
Due to this, the hermitian $M_l M_l^\dagger$ combination is not going to be diagonal at leading order as it was for the quarks.
We assign the $SU(2)$ singlet charged leptons $e^c$, $\mu^c$, $\tau^c$ as $1_{00}$, $1_{00}$ and $1_{02}$ (note that these are not the singlets used in \cite{Ma:2013xqa}). The leading invariants are generically of the type $H\left[ (L \phi^\dagger) l^c_i\right]$, e.g.
\begin{align}
  & H \left[y_3 (L \phi^\dagger)_{01} \tau^c + y_2 (L \phi^\dagger)_{02} \mu^c (\theta^2) + y_1 (L \phi^\dagger)_{00} e^c (\theta^3) \right] .
  \label{eqn:LInvariants}
\end{align}
The corresponding mass matrix is
\begin{align}
  M_{l}  &=  
  \begin{pmatrix}
    y_{1}\omega^{2}  & y_{2} & y_{3} \\
    y_{1} & y_{2} & y_{3}\omega^{2} \\
    y_{1} & y_{2}\omega^{2} & y_{3}
  \end{pmatrix} ,
  \label{eqn:MLdef}
\end{align}
where we have reabsorbed the VEVs into the $y_i$. The hermitian combination
\begin{align}
  M_{l}M_{l}^{\dagger}  &=  
  \begin{pmatrix}
    y_{1}^{2} + y_{2}^{2} + y_{3}^{2}  & y_{1}^{2}\omega^{2} + y_{2}^{2} + y_{3}^{2}\omega & y_{1}^{2}\omega^{2} + y_{2}^{2}\omega + y_{3}^{2} \\
    y_{1}^{2}\omega^{2} + y_{2}^{2} + y_{3}^{2}\omega & y_{1}^{2} + y_{2}^{2} + y_{3}^{2} & y_{1}^{2} + y_{2}^{2}\omega + y_{3}^{2}\omega^{2} \\
    y_{1}^{2}\omega^{2} + y_{2}^{2}\omega + y_{3}^{2} & y_{1}^{2} + y_{2}^{2}\omega + y_{3}^{2}\omega^{2} & y_{1}^{2} + y_{2}^{2} + y_{3}^{2}
  \end{pmatrix} ,
  \label{eqn:MLtmesMLDagger}
\end{align}
is diagonalized by
\begin{align}
  V_{l}  &= \frac{1}{\sqrt{3}} 
  \begin{pmatrix}
    \omega^{2}  & \omega & 1 \\
    1 & \omega & \omega^{2} \\
    1 & 1 & 1
  \end{pmatrix} .
  \label{eqn:VLdef}
\end{align}
$V_l$ has no free parameters and the parameters $y_i$ are fixed by the charged lepton masses.

As a brief aside that illustrates to some extent the difficulty in obtaining fermion masses and mixing in this framework, it is interesting to consider a couple of scenarios.
With $(L L \phi)$ invariants we readily encounter promising structures where the mixing in the neutrino sector is approximately tribimaximal (TB).
Had the charged leptons been diagonal, these solutions would be viable, but with the $V_l$ in eq.(\ref{eqn:VLdef}) the mixing gets spoiled.
One would need to contract $L$ with a different triplet field aligned as $(1,0,0)$ in the charged lepton sector to make these neutrino solutions viable.
Conversely, had we assigned instead the charged lepton $SU(2)$ singlets generations as a $\Delta(27)$ triplet and $L_1$, $L_2$, $L_3$ as $1_{00}$, $1_{01}$, $1_{02}$, we could have a leading order neutrino mass matrix where the only non-zero entries are $11$, $23$ and $32$. This $\mu - \tau$ interchange could be a promising starting point to obtain viable mixing in combination with the $V_l$ in eq.(\ref{eqn:VLdef}), but if the $L_i$ are singlets we obtain diagonal charged lepton masses (in analogy with the leading quark structures). Once again, one would need a field aligned in the $(1,0,0)$ direction, this time contracted with the $LL$ operator to obtain the $\mu - \tau$ interchange structure for the neutrinos together with eq.(\ref{eqn:VLdef}).
In both cases the symmetry would be broken to different directions in each sector, the traditional strategy to generate (near) TB mixing with a discrete group.
But because we want to preserve GCPV, we avoid introducing additional triplet scalars.

We proceed now with the detailed discussion of invariants of type $H^\dagger H^\dagger (L L \phi)$. There are two symmetrical $3_{02}$ constructed from $L L$ that contract with $\phi$, each generating one of each $\Delta(27)$ singlet.
The two distinct $H^\dagger H^\dagger (L L \phi)_{00}$ invariants are not enough for viable masses and mixing, so we consider in general that there are auxiliary (or spurion) fields $\xi$, $\xi'$ and $\xi''$ transforming as $1_{00}$, $1_{01}$ and $1_{02}$, all sharing the opposite FN charge of $H^\dagger H^\dagger (L L \phi)$ (these spurions can be constructed from combinations of the physical fields $\varphi$ and $\theta$, see section \ref{sec:model}).
Then we have two invariants for each auxiliary field:
\begin{align}
  &H^\dagger H^\dagger \xi \left[ z_1 (L_i L_i \phi_i)_{00}  + z_4 (L_i L_j \phi_k)_{00} \right]\\
  &H^\dagger H^\dagger \xi' \left[ z_2 (L_i L_i \phi_i)_{02} + z_5 (L_i L_j \phi_k)_{02} \right]\\
  &H^\dagger H^\dagger \xi'' \left[ z_3 (L_i L_i \phi_i)_{01} + z_6 (L_i L_j \phi_k)_{01} \right] ,
  \label{eqn:NuInvariants}
\end{align}
where the $ijk$ invariant denotes the symmetric contraction, e.g.
\begin{align}
  (L_i L_j \phi_k)_{00} &=  L_2 L_3 \phi_1 + L_3 L_1 \phi_2 + L_1 L_2 \phi_3
  \label{eqn:NuInvariants2}
\end{align}
\begin{align}
  (L_i L_j \phi_k)_{02} &=  L_2 L_3 \phi_2 + L_3 L_1 \phi_3 + L_1 L_2 \phi_1
  \label{eqn:NuInvariants3}
\end{align}

The invariants correspond to mass structures
\begin{align}
  M_{\xi}  &=  
  \begin{pmatrix}
    z_{1}\omega^{2}  & z_{4} & z_{4} \\
    z_{4} & z_{1} & z_{4}\omega^{2} \\
    z_{4} & z_{4}\omega^{2} & z_{1}
  \end{pmatrix} ,
  \label{eqn:Mxi}
\end{align}
\begin{align}
  M_{\xi'}  &=  
  \begin{pmatrix}
    z_{2} & z_{5} \omega^{2} & z_{5} \\
    z_{5} \omega^{2} & z_{2} & z_{5} \\
    z_{5} & z_{5} & z_{2} \omega^{2}
  \end{pmatrix} ,
  \label{eqn:Mxip}
\end{align}
\begin{align}
  M_{\xi''}  &=  
  \begin{pmatrix}
    z_{3} & z_{6} & z_{6} \omega^{2} \\
    z_{6}  & z_{3}\omega^{2} & z_{6} \\
    z_{6} \omega^{2} & z_{6} & z_{3} 
  \end{pmatrix} .
  \label{eqn:Mxipp}
\end{align}

In a phenomenological scan, we have 6 free $z_i$ parameters and need to obtain the mass splittings and mixing angles (with $V_l$ taken into account).
While the mass structures in eqs.(\ref{eqn:Mxi}-\ref{eqn:Mxipp}) are fairly constraining, we found regions of parameter space with viable squared mass splittings and mixing angles in the 3-$\sigma$ ranges of \cite{GonzalezGarcia:2012sz}. Table \ref{ta:six} contains one example for each hierarchy, the inverted (IH) and the normal (NH).


We now consider also invariants of the type $H^\dagger H^\dagger (L \phi^\dagger) (L \phi^\dagger)$, namely
\begin{align}
  &H^\dagger H^\dagger \left[ A (L \phi^\dagger)_{00} (L \phi^\dagger)_{00}  + B (L \phi^\dagger)_{01} (L \phi^\dagger)_{02} \right]
  \label{eqn:NuInvariants4}
\end{align}
With the additional parameters $A$, $B$, minimal models only require contributions from two of the auxiliary fields.
We scanned the three classes of models and obtained for each large regions of viable parameter space. Tables \ref{xixip}, \ref{xixipp} and \ref{xipxipp} have some examples.


In order to have an idea of the fine-tuning, we have relied on the procedure discussed in \cite{Altarelli:2010at}: we use $d_{FT}$ as a quantitative measure of the fine-tuning, a dimensionless quantity defined as the sum of the absolute values of ratios between all parameters and respective errors, where these errors are themselves defined as the deviation in that parameter that leads to an increase of $\chi^2$ by 1 (while the other parameters remain at their fitted values). In \cite{Altarelli:2010at}, another similar quantity $d_{Data}$ is introduced, defined simply the sum of the absolute values of ratios between the data and respective errors - we also present this number in the appendix.

\section{Specific models \label{sec:model}}

In this section we present an example of how to build concrete models by assigning specific FN charges to the lepton sector in a way that is consistent with the desired quark terms.
We leave $H$ neutral for simplicity and retain the charges of $\varphi$, $\theta$.
The charges of $u^c$, $d^c$ are adjusted to cancel the charge of $\phi$, such that the combinations $H^\dagger Q u^c \phi$ and $H Q d^c \phi$ are overall neutral under $U(1)_F$.
The charges of the different $Q_i$ generations select different combinations of $\varphi$ and $\theta$ producing exactly the same FN suppressions as in \cite{Varzielas:2013sla}.

An important distinction is whether any hierarchy in the effective couplings $z_i$ is between terms involving different $\xi$, $\xi'$ and $\xi''$ or not: a large hierarchy within the two invariants involving the same spurion is unnatural,
as the invariants correspond to the same physical field combination. An hierarchy between $A$, $B$ is similarly not desirable. We first selected solutions where there are only natural couplings (avoiding the issue described above), then chose two particularly interesting types where the FN charges look simpler. These cases have $z_1$, $z_4$ about one order of magnitude larger than $A$,$B$, $z_2$, $z_5$, $z_3$, $z_6$, illustrated in table \ref{Numodels12}. $d_{FT}$ is also shown for these examples.

We have one type of model with
\begin{align}
  &H^\dagger H^\dagger (\theta^3)^\dagger \left[ A (L \phi^\dagger)_{00} (L \phi^\dagger)_{00}  + B (L \phi^\dagger)_{01} (L \phi^\dagger)_{02} \right]\\
  +&H^\dagger H^\dagger \theta^3 \left[ z_1 (L_i L_i \phi_i)_{00} + z_4 (L_i L_j \phi_k)_{00} \right]\\
  +&H^\dagger H^\dagger \theta^2 \varphi^2 \left[ z_2 (L_i L_i \phi_i)_{02} + z_5 (L_i L_j \phi_k)_{02} \right] .
  \label{eqn:NuModel1}
\end{align}
Another type of model has
\begin{align}
  &H^\dagger H^\dagger \theta^3 \left[ A (L \phi^\dagger)_{00} (L \phi^\dagger)_{00}  + B (L \phi^\dagger)_{01} (L \phi^\dagger)_{02} \right]\\
  +&H^\dagger H^\dagger (\theta^3)^\dagger \left[ z_1 (L_i L_i \phi_i)_{00} + z_4 (L_i L_j \phi_k)_{00} \right]\\
  +&H^\dagger H^\dagger (\theta^2 \varphi^2)^\dagger \left[ z_3 (L_i L_i \phi_i)_{01} + z_6 (L_i L_j \phi_k)_{01} \right] .
  \label{eqn:NuModel2}
\end{align}

At this stage it is clear that we can identify the auxiliary fields $\xi$, $\xi'$ or $\xi''$ as combinations of $\varphi$ and $\theta$.
In any case, $z_1$, $z_4$ appear in invariants with 8 field insertions compared to the other neutrino mass invariants at 9 field insertions, so the hierarchy between them is natural
as it is accounted for by the FN assignments listed in Table {\ref{ta:imp}}.
The charges of $L$ and $\phi$ are selected to make $H^\dagger H^\dagger (L \phi^\dagger) (L \phi^\dagger)$ invariant either with $\theta^3$ or $(\theta^3)^\dagger$ (9 field insertions for the $A$, $B$ parameters). 
\footnote{The other two combinations where $H^\dagger H^\dagger (L \phi^\dagger) (L \phi^\dagger)$ has the same overall FN charge as $H^\dagger H^\dagger (L L \phi)$ corresponds to the charge assignment $p=0$, which is not compatible with our requirement for the scalar potential not to have he term $\phi^3$.}

\begin{table}
  \centering
  \begin{tabular}{|c||cc|ccc|}
    \hline
    & $L$ & $\phi$ &  $e^c$ & $\mu^c$ & $\tau^c$  \\
    \hline
$1_{00}$+$1_{01}$  & -1  & -4 &  3  & 1  & -3  \\
$1_{00}$+$1_{02}$  & 1 & 4 &  9 & 7 & 3 \\    
    \hline
  \end{tabular}
  \caption{Specific $U(1)_F$ charges for two sample models that are considered natural in terms of hierarchies. \label{ta:imp}}
\end{table}

\section{Scalar potential \label{sec:scalar}}

We consider now the full renormalisable scalar potential. As long as problematic FN charge assignments for $\phi$ (such as $p=0$) are carefully avoided,
there are no renormalisable terms that are cubic in $\phi$.
Similarly, the FN assignments of $\varphi$ and $\theta$ guarantee the absence of terms such as $(\phi^\dagger \phi)_{01} \theta$, $(\phi^\dagger \phi)_{02} \theta^\dagger$ or any other phase-dependent $\phi$ invariants that would spoil GCPV as in \cite{Varzielas:2013sla}. The potential is then
\begin{align}
&V(H, \phi, \varphi, \theta) = m_H^2 H H^\dagger + m_\varphi^2 \varphi \varphi^\dagger + m_\theta^2 \theta \theta^\dagger\\
&+ \lambda_H (H H^\dagger)^2 + \lambda_\varphi (\varphi\varphi^\dagger)^2 +\lambda_\theta (\theta \theta^\dagger)^2+ \lambda_{\varphi \theta}(\varphi\varphi^\dagger)(\theta \theta^\dagger)  \\
&+ \left(\lambda_{\varphi H} \varphi \varphi^\dagger +  \lambda_{\theta H} \theta \theta^\dagger \right) \left( H H^\dagger \right)  \\
&+ m_\phi^2 \left[ \phi_i \phi_i^\dagger \right]
+\lambda_1 \left[ (\phi_i \phi_i^\dagger)^2
\right]+
\lambda_2 \left( \phi_1 \phi_1^\dagger \phi_2 \phi_2^\dagger + \phi_2 \phi_2^\dagger \phi_3 \phi_3^\dagger + \phi_3 \phi_3^\dagger \phi_1 \phi_1^\dagger \right)\\
&+\lambda_3 \left( \phi_1 \phi_2^\dagger \phi_1 \phi_3^\dagger + \phi_2 \phi_3^\dagger \phi_2 \phi_1^\dagger +\phi_3 \phi_1^\dagger \phi_3 \phi_2^\dagger +\text{h.c.} \right)\\
&+\left(\lambda_{H \phi} H H^\dagger + \lambda_{\varphi \phi} \varphi \varphi^\dagger +  \lambda_{\theta \phi} \theta \theta^\dagger \right) \left[ \phi_i \phi_i^\dagger \right]
\,.
\end{align}
The notable property of this potential that leads to GCPV is precisely that there is only one invariant that depends on complex phases, and when $\lambda_3$ is positive the VEV for $\phi$ is of the form shown eq.(\ref{eq:VEV}).
The magnitudes of the VEVs of $H$, $\phi$, $\varphi$ and $\theta$ are controlled by the various mass terms and quartic couplings and can readily be distinct.
It is only when all the scalars acquire their respective VEVs that the electroweak symmetry, $\Delta(27)$ and the FN symmetry are broken, and through this breaking the three generations of fermions acquire their masses.

\section{Summary}

For different frameworks, we performed an extensive phenomenological scan and found several regions of parameter space where viable lepton masses and mixing are obtained.
We then constructed specific models for some promising examples where the model parameters are natural.
These models are entirely consistent with the previously found solution for the quark sector, so we have formulated an existence proof of models of quark and lepton masses and mixing that feature geometrical CP violation.

\acknowledgments

IdMV was supported by DFG grant PA 803/6-1 and by the Swiss National Science Foundation.\\
DP was supported by DFG grant PA 803/5-1.

\newpage

\appendix

\section{Tables containing numerical results \label{A}}

This appendix contains tables with sample numerical values for the parameters in different classes of models.
For each class we found many sets of values (hits) that lead to viable respective mixing angles and neutrino squared mass splittings (not displayed),
within the $3 \sigma$ experimentally allowed ranges.

Table \ref{ta:six} corresponds to models with the full set of 6 $z_i$ parameters,
whereas tables \ref{xixip}, \ref{xixipp} and \ref{xipxipp} have some examples for each of the three classes with $A$, $B$.
Table \ref{Numodels12} has examples where $z_1$, $z_4$ is larger by one order of magnitude.

For comparison, $d_{Data} = 39.4773$. As can be seen in the tables below, for the hits we display, $d_{FT}$ is no more than one or two orders of magnitude higher than $d_{Data}$.

\begin{table}[h]
  \centering
\tiny{
  \begin{tabular}{cccccccc}
    $z_1$  & $z_2$ & $z_3$  & $z_4$  & $z_5$  & $z_6$  & $\chi^{2}$ & $d_{FT}$ \\
    \hline\\
    -0.00554161	&	-0.00340302	&	-0.00227104	&	-0.0141038	&	0.0175277	&	0.0170266	&	0.46992	&	640.595	\\
    0.00967825	&	-0.0118758	&	-0.00670678	&	-0.0160151	&	-0.00629062	&	-0.00653824	&	0.187884	&	246.89	\\
  \end{tabular}
   \caption{Sample hits for the 6 $z_i$ model. First row is for IH, second row for NH. \label{ta:six}}
}
\end{table}

\begin{table}[hp]
  \centering
\tiny{
  \begin{tabular}{cccccccc}
    $A$  &  $B$ & $z_{1}$  & $z_{4}$  & $z_{2}$  & $z_{5}$  & $\chi^{2}$ & $d_{FT}$ \\
    \hline
    -0.00353652	&	0.00107432	&	-0.0524306	&	-0.00585345	&	-0.00696862	&	0.0118005	&	1.161	&	516.293	\\
    0.0109313	&	-0.0215866	&	0.0172491	&	0.0154776	&	-0.00496799	&	-0.00163566	&	0.247987	&	1480.21	\\
    0.00907931	&	-0.0256511	&	-0.00227895	&	0.0142952	&	0.00284323	&	0.00881012	&	0.389833	&	1074.57	\\
\end{tabular}
   \caption{Sample hits for the contractions $\xi$ and $\xi'$ in the $A$,$B$ class of models. \label{xixip}}
}
\end{table}

\begin{table}[hp]
  \centering
\tiny{
  \begin{tabular}{cccccccc}
    $A$  &  $B$ & $z_{1}$  & $z_{4}$  & $z_{3}$  & $z_{6}$  & $\chi^{2}$ & $d_{FT}$ \\
    \hline\\
    -0.00941157	&	0.00922199	&	0.0190035	&	-0.0103075	&	-0.0277012	&	-0.045596	&	0.791304	&	3546.25	\\
    -0.0141045	&	0.0349938	&	-0.00117646	&	-0.00713608	&	-0.00700718	&	0.00389799	&	0.277284	&	2234.42	\\
  \end{tabular}
   \caption{Sample hits for the contractions $\xi$ and $\xi''$ in the $A$,$B$ class of models. \label{xixipp}}
}
\end{table}

\begin{table}[h]
  \centering
\tiny{
  \begin{tabular}{cccccccc}
    $A$  &  $B$ & $z_{2}$  & $z_{5}$  & $z_{3}$  & $z_{6}$  & $\chi^{2}$ & $d_{FT}$ \\  
    \hline\\
    0.0099768	&	-0.0387091	&	-0.00543329	&	0.00613746	&	-0.0101139	&	0.0284193	&	0.607939	&	4035.83	\\
    0.00438667	&	-0.0049329	&	0.00337298	&	-0.00261386 &	0.0616697	&	-0.0127218	&	0.132725	&	1301.91	\\
    0.00476606	&	-0.00529681	&	0.0608295	&	-0.0129415	&	0.00381149	&	-0.00135987	&	0.915034	&	1269.28	\\
	\end{tabular}
	\caption{Sample hits for the contractions $\xi'$ and $\xi''$ in the $A$,$B$ class of models. \label{xipxipp}}
}
\end{table}

\begin{table}[h]
 \centering
\tiny{
 \begin{tabular}{cccccccc}
   $A$  &  $B$ & $z_1$  & $z_4$  & $z_2$  & $z_5$  & $\chi^{2}$ & $d_{FT}$ \\ \hline
	0.00245874	&	-0.00750093	&	0.0561966	&	0.0143339	&	0.0045167	&	-0.00187831	&	0.901573	&	725.276 \\ 
	\\
   $A$  &  $B$ & $z_1$  & $z_4$  & $z_3$  & $z_6$  & $\chi^{2}$ & $d_{FT}$ \\ \hline
	0.00153913	&	-0.00573201	&	0.0427374	&	0.0325508	&	0.00544977	&	-0.0013738	&	0.249697	&	640.799	\\ 
 \end{tabular}
  \caption{
  A sample hit for the contractions $\xi$ and $\xi'$ (top) and the $\xi$ and $\xi''$ (bottom) in the $A$,$B$ class of models matching the natural hierarchies associated with the FN charges listed in Table \ref{ta:imp}.
    \label{Numodels12}}
}
\end{table}

\newpage

\bibliography{v2}

\begin{thebibliography}{19}%
\makeatletter
\providecommand \@ifxundefined [1]{%
 \@ifx{#1\undefined}
}%
\providecommand \@ifnum [1]{%
 \ifnum #1\expandafter \@firstoftwo
 \else \expandafter \@secondoftwo
 \fi
}%
\providecommand \@ifx [1]{%
 \ifx #1\expandafter \@firstoftwo
 \else \expandafter \@secondoftwo
 \fi
}%
\providecommand \natexlab [1]{#1}%
\providecommand \enquote  [1]{``#1''}%
\providecommand \bibnamefont  [1]{#1}%
\providecommand \bibfnamefont [1]{#1}%
\providecommand \citenamefont [1]{#1}%
\providecommand \href@noop [0]{\@secondoftwo}%
\providecommand \href [0]{\begingroup \@sanitize@url \@href}%
\providecommand \@href[1]{\@@startlink{#1}\@@href}%
\providecommand \@@href[1]{\endgroup#1\@@endlink}%
\providecommand \@sanitize@url [0]{\catcode `\\12\catcode `\$12\catcode
  `\&12\catcode `\#12\catcode `\^12\catcode `\_12\catcode `\%12\relax}%
\providecommand \@@startlink[1]{}%
\providecommand \@@endlink[0]{}%
\providecommand \url  [0]{\begingroup\@sanitize@url \@url }%
\providecommand \@url [1]{\endgroup\@href {#1}{\urlprefix }}%
\providecommand \urlprefix  [0]{URL }%
\providecommand \Eprint [0]{\href }%
\providecommand \doibase [0]{http://dx.doi.org/}%
\providecommand \selectlanguage [0]{\@gobble}%
\providecommand \bibinfo  [0]{\@secondoftwo}%
\providecommand \bibfield  [0]{\@secondoftwo}%
\providecommand \translation [1]{[#1]}%
\providecommand \BibitemOpen [0]{}%
\providecommand \bibitemStop [0]{}%
\providecommand \bibitemNoStop [0]{.\EOS\space}%
\providecommand \EOS [0]{\spacefactor3000\relax}%
\providecommand \BibitemShut  [1]{\csname bibitem#1\endcsname}%
\let\auto@bib@innerbib\@empty
\bibitem [{\citenamefont {Branco}\ \emph {et~al.}(1984)\citenamefont {Branco},
  \citenamefont {Gerard},\ and\ \citenamefont {Grimus}}]{Branco:1983tn}%
  \BibitemOpen
  \bibfield  {author} {\bibinfo {author} {\bibfnamefont {G.}~\bibnamefont
  {Branco}}, \bibinfo {author} {\bibfnamefont {J.}~\bibnamefont {Gerard}}, \
  and\ \bibinfo {author} {\bibfnamefont {W.}~\bibnamefont {Grimus}},\ }\href
  {\doibase 10.1016/0370-2693(84)92024-0} {\bibfield  {journal} {\bibinfo
  {journal} {Phys.Lett.}\ }\textbf {\bibinfo {volume} {B136}},\ \bibinfo
  {pages} {383} (\bibinfo {year} {1984})}\BibitemShut {NoStop}%
\bibitem [{\citenamefont {de~Medeiros~Varzielas}\ and\ \citenamefont
  {Emmanuel-Costa}(2011)}]{deMedeirosVarzielas:2011zw}%
  \BibitemOpen
  \bibfield  {author} {\bibinfo {author} {\bibfnamefont {I.}~\bibnamefont
  {de~Medeiros~Varzielas}}\ and\ \bibinfo {author} {\bibfnamefont
  {D.}~\bibnamefont {Emmanuel-Costa}},\ }\href {\doibase
  10.1103/PhysRevD.84.117901} {\bibfield  {journal} {\bibinfo  {journal}
  {Phys.Rev.}\ }\textbf {\bibinfo {volume} {D84}},\ \bibinfo {pages} {117901}
  (\bibinfo {year} {2011})},\ \Eprint {http://arxiv.org/abs/1106.5477}
  {arXiv:1106.5477 [hep-ph]} \BibitemShut {NoStop}%
\bibitem [{\citenamefont {de~Medeiros~Varzielas}\ \emph
  {et~al.}(2012)\citenamefont {de~Medeiros~Varzielas}, \citenamefont
  {Emmanuel-Costa},\ and\ \citenamefont {Leser}}]{Varzielas:2012nn}%
  \BibitemOpen
  \bibfield  {author} {\bibinfo {author} {\bibfnamefont {I.}~\bibnamefont
  {de~Medeiros~Varzielas}}, \bibinfo {author} {\bibfnamefont {D.}~\bibnamefont
  {Emmanuel-Costa}}, \ and\ \bibinfo {author} {\bibfnamefont {P.}~\bibnamefont
  {Leser}},\ }\href {\doibase 10.1016/j.physletb.2012.08.008} {\bibfield
  {journal} {\bibinfo  {journal} {Phys.Lett.}\ }\textbf {\bibinfo {volume}
  {B716}},\ \bibinfo {pages} {193} (\bibinfo {year} {2012})},\ \Eprint
  {http://arxiv.org/abs/1204.3633} {arXiv:1204.3633 [hep-ph]} \BibitemShut
  {NoStop}%
\bibitem [{\citenamefont {de~Medeiros~Varzielas}(2012)}]{Varzielas:2012pd}%
  \BibitemOpen
  \bibfield  {author} {\bibinfo {author} {\bibfnamefont {I.}~\bibnamefont
  {de~Medeiros~Varzielas}},\ }\href {\doibase 10.1007/JHEP08(2012)055}
  {\bibfield  {journal} {\bibinfo  {journal} {JHEP}\ }\textbf {\bibinfo
  {volume} {1208}},\ \bibinfo {pages} {055} (\bibinfo {year} {2012})},\ \Eprint
  {http://arxiv.org/abs/1205.3780} {arXiv:1205.3780 [hep-ph]} \BibitemShut
  {NoStop}%
\bibitem [{\citenamefont {Bhattacharyya}\ \emph {et~al.}(2012)\citenamefont
  {Bhattacharyya}, \citenamefont {de~Medeiros~Varzielas},\ and\ \citenamefont
  {Leser}}]{Bhattacharyya:2012pi}%
  \BibitemOpen
  \bibfield  {author} {\bibinfo {author} {\bibfnamefont {G.}~\bibnamefont
  {Bhattacharyya}}, \bibinfo {author} {\bibfnamefont {I.}~\bibnamefont
  {de~Medeiros~Varzielas}}, \ and\ \bibinfo {author} {\bibfnamefont
  {P.}~\bibnamefont {Leser}},\ }\href {\doibase 10.1103/PhysRevLett.109.241603}
  {\bibfield  {journal} {\bibinfo  {journal} {Phys.Rev.Lett.}\ }\textbf
  {\bibinfo {volume} {109}},\ \bibinfo {pages} {241603} (\bibinfo {year}
  {2012})},\ \Eprint {http://arxiv.org/abs/1210.0545} {arXiv:1210.0545
  [hep-ph]} \BibitemShut {NoStop}%
\bibitem [{\citenamefont {Ivanov}\ and\ \citenamefont
  {Lavoura}(2013)}]{Ivanov:2013nla}%
  \BibitemOpen
  \bibfield  {author} {\bibinfo {author} {\bibfnamefont {I.}~\bibnamefont
  {Ivanov}}\ and\ \bibinfo {author} {\bibfnamefont {L.}~\bibnamefont
  {Lavoura}},\ }\href {\doibase 10.1140/epjc/s10052-013-2416-8} {\bibfield
  {journal} {\bibinfo  {journal} {Eur.Phys.J.}\ }\textbf {\bibinfo {volume}
  {C73}},\ \bibinfo {pages} {2416} (\bibinfo {year} {2013})},\ \Eprint
  {http://arxiv.org/abs/1302.3656} {arXiv:1302.3656 [hep-ph]} \BibitemShut
  {NoStop}%
\bibitem [{\citenamefont {de~Medeiros~Varzielas}(2013)}]{Varzielas:2013zbp}%
  \BibitemOpen
  \bibfield  {author} {\bibinfo {author} {\bibfnamefont {I.}~\bibnamefont
  {de~Medeiros~Varzielas}},\ }\href@noop {} {\  (\bibinfo {year} {2013})},\
  \Eprint {http://arxiv.org/abs/1302.3991} {arXiv:1302.3991 [hep-ph]}
  \BibitemShut {NoStop}%
\bibitem [{\citenamefont {Ma}(2013)}]{Ma:2013xqa}%
  \BibitemOpen
  \bibfield  {author} {\bibinfo {author} {\bibfnamefont {E.}~\bibnamefont
  {Ma}},\ }\href {\doibase 10.1016/j.physletb.2013.05.011} {\bibfield
  {journal} {\bibinfo  {journal} {Phys.Lett.}\ }\textbf {\bibinfo {volume}
  {B723}},\ \bibinfo {pages} {161} (\bibinfo {year} {2013})},\ \Eprint
  {http://arxiv.org/abs/1304.1603} {arXiv:1304.1603 [hep-ph]} \BibitemShut
  {NoStop}%
\bibitem [{\citenamefont {de~Medeiros~Varzielas}\ and\ \citenamefont
  {Pidt}(2013)}]{Varzielas:2013sla}%
  \BibitemOpen
  \bibfield  {author} {\bibinfo {author} {\bibfnamefont {I.}~\bibnamefont
  {de~Medeiros~Varzielas}}\ and\ \bibinfo {author} {\bibfnamefont
  {D.}~\bibnamefont {Pidt}},\ }\href@noop {} {\  (\bibinfo {year} {2013})},\
  \Eprint {http://arxiv.org/abs/1307.0711} {arXiv:1307.0711 [hep-ph]}
  \BibitemShut {NoStop}%
\bibitem [{\citenamefont {Feruglio}\ \emph {et~al.}(2013)\citenamefont
  {Feruglio}, \citenamefont {Hagedorn},\ and\ \citenamefont
  {Ziegler}}]{Feruglio:2012cw}%
  \BibitemOpen
  \bibfield  {author} {\bibinfo {author} {\bibfnamefont {F.}~\bibnamefont
  {Feruglio}}, \bibinfo {author} {\bibfnamefont {C.}~\bibnamefont {Hagedorn}},
  \ and\ \bibinfo {author} {\bibfnamefont {R.}~\bibnamefont {Ziegler}},\ }\href
  {\doibase 10.1007/JHEP07(2013)027} {\bibfield  {journal} {\bibinfo  {journal}
  {JHEP}\ }\textbf {\bibinfo {volume} {1307}},\ \bibinfo {pages} {027}
  (\bibinfo {year} {2013})},\ \Eprint {http://arxiv.org/abs/1211.5560}
  {arXiv:1211.5560 [hep-ph]} \BibitemShut {NoStop}%
\bibitem [{\citenamefont {Holthausen}\ \emph {et~al.}(2013)\citenamefont
  {Holthausen}, \citenamefont {Lindner},\ and\ \citenamefont
  {Schmidt}}]{Holthausen:2012dk}%
  \BibitemOpen
  \bibfield  {author} {\bibinfo {author} {\bibfnamefont {M.}~\bibnamefont
  {Holthausen}}, \bibinfo {author} {\bibfnamefont {M.}~\bibnamefont {Lindner}},
  \ and\ \bibinfo {author} {\bibfnamefont {M.~A.}\ \bibnamefont {Schmidt}},\
  }\href {\doibase 10.1007/JHEP04(2013)122} {\bibfield  {journal} {\bibinfo
  {journal} {JHEP}\ }\textbf {\bibinfo {volume} {1304}},\ \bibinfo {pages}
  {122} (\bibinfo {year} {2013})},\ \Eprint {http://arxiv.org/abs/1211.6953}
  {arXiv:1211.6953 [hep-ph]} \BibitemShut {NoStop}%
\bibitem [{\citenamefont {Froggatt}\ and\ \citenamefont
  {Nielsen}(1979)}]{Froggatt:1978nt}%
  \BibitemOpen
  \bibfield  {author} {\bibinfo {author} {\bibfnamefont {C.}~\bibnamefont
  {Froggatt}}\ and\ \bibinfo {author} {\bibfnamefont {H.~B.}\ \bibnamefont
  {Nielsen}},\ }\href {\doibase 10.1016/0550-3213(79)90316-X} {\bibfield
  {journal} {\bibinfo  {journal} {Nucl.Phys.}\ }\textbf {\bibinfo {volume}
  {B147}},\ \bibinfo {pages} {277} (\bibinfo {year} {1979})}\BibitemShut
  {NoStop}%
\bibitem [{\citenamefont {de~Medeiros~Varzielas}\ and\ \citenamefont
  {Merlo}(2011)}]{Varzielas:2010mp}%
  \BibitemOpen
  \bibfield  {author} {\bibinfo {author} {\bibfnamefont {I.}~\bibnamefont
  {de~Medeiros~Varzielas}}\ and\ \bibinfo {author} {\bibfnamefont
  {L.}~\bibnamefont {Merlo}},\ }\href {\doibase 10.1007/JHEP02(2011)062}
  {\bibfield  {journal} {\bibinfo  {journal} {JHEP}\ }\textbf {\bibinfo
  {volume} {1102}},\ \bibinfo {pages} {062} (\bibinfo {year} {2011})},\ \Eprint
  {http://arxiv.org/abs/1011.6662} {arXiv:1011.6662 [hep-ph]} \BibitemShut
  {NoStop}%
\bibitem [{\citenamefont {Medeiros~Varzielas}\ and\ \citenamefont
  {Pidt}(2013)}]{Varzielas:2012ai}%
  \BibitemOpen
  \bibfield  {author} {\bibinfo {author} {\bibfnamefont {I.}~\bibnamefont
  {Medeiros~Varzielas}}\ and\ \bibinfo {author} {\bibfnamefont
  {D.}~\bibnamefont {Pidt}},\ }\href {\doibase 10.1007/JHEP03(2013)065}
  {\bibfield  {journal} {\bibinfo  {journal} {JHEP}\ }\textbf {\bibinfo
  {volume} {1303}},\ \bibinfo {pages} {065} (\bibinfo {year} {2013})},\ \Eprint
  {http://arxiv.org/abs/1211.5370} {arXiv:1211.5370 [hep-ph]} \BibitemShut
  {NoStop}%
\bibitem [{\citenamefont {Luhn}(2013)}]{Luhn:2013lkn}%
  \BibitemOpen
  \bibfield  {author} {\bibinfo {author} {\bibfnamefont {C.}~\bibnamefont
  {Luhn}},\ }\href@noop {} {\  (\bibinfo {year} {2013})},\ \Eprint
  {http://arxiv.org/abs/1306.2358} {arXiv:1306.2358 [hep-ph]} \BibitemShut
  {NoStop}%
\bibitem [{\citenamefont {Antusch}\ \emph {et~al.}(2013)\citenamefont
  {Antusch}, \citenamefont {Holthausen}, \citenamefont {Schmidt},\ and\
  \citenamefont {Spinrath}}]{Antusch:2013rla}%
  \BibitemOpen
  \bibfield  {author} {\bibinfo {author} {\bibfnamefont {S.}~\bibnamefont
  {Antusch}}, \bibinfo {author} {\bibfnamefont {M.}~\bibnamefont {Holthausen}},
  \bibinfo {author} {\bibfnamefont {M.~A.}\ \bibnamefont {Schmidt}}, \ and\
  \bibinfo {author} {\bibfnamefont {M.}~\bibnamefont {Spinrath}},\ }\href@noop
  {} {\  (\bibinfo {year} {2013})},\ \Eprint {http://arxiv.org/abs/1307.0710}
  {arXiv:1307.0710 [hep-ph]} \BibitemShut {NoStop}%
\bibitem [{\citenamefont {Ding}\ \emph {et~al.}(2013)\citenamefont {Ding},
  \citenamefont {King},\ and\ \citenamefont {Stuart}}]{Ding:2013bpa}%
  \BibitemOpen
  \bibfield  {author} {\bibinfo {author} {\bibfnamefont {G.-J.}\ \bibnamefont
  {Ding}}, \bibinfo {author} {\bibfnamefont {S.~F.}\ \bibnamefont {King}}, \
  and\ \bibinfo {author} {\bibfnamefont {A.~J.}\ \bibnamefont {Stuart}},\
  }\href@noop {} {\  (\bibinfo {year} {2013})},\ \Eprint
  {http://arxiv.org/abs/1307.4212} {arXiv:1307.4212 [hep-ph]} \BibitemShut
  {NoStop}%
\bibitem [{\citenamefont {Gonzalez-Garcia}\ \emph {et~al.}(2012)\citenamefont
  {Gonzalez-Garcia}, \citenamefont {Maltoni}, \citenamefont {Salvado},\ and\
  \citenamefont {Schwetz}}]{GonzalezGarcia:2012sz}%
  \BibitemOpen
  \bibfield  {author} {\bibinfo {author} {\bibfnamefont {M.}~\bibnamefont
  {Gonzalez-Garcia}}, \bibinfo {author} {\bibfnamefont {M.}~\bibnamefont
  {Maltoni}}, \bibinfo {author} {\bibfnamefont {J.}~\bibnamefont {Salvado}}, \
  and\ \bibinfo {author} {\bibfnamefont {T.}~\bibnamefont {Schwetz}},\ }\href
  {\doibase 10.1007/JHEP12(2012)123} {\bibfield  {journal} {\bibinfo  {journal}
  {JHEP}\ }\textbf {\bibinfo {volume} {1212}},\ \bibinfo {pages} {123}
  (\bibinfo {year} {2012})},\ \Eprint {http://arxiv.org/abs/1209.3023}
  {arXiv:1209.3023 [hep-ph]} \BibitemShut {NoStop}%
\bibitem [{\citenamefont {Altarelli}\ and\ \citenamefont
  {Blankenburg}(2011)}]{Altarelli:2010at}%
  \BibitemOpen
  \bibfield  {author} {\bibinfo {author} {\bibfnamefont {G.}~\bibnamefont
  {Altarelli}}\ and\ \bibinfo {author} {\bibfnamefont {G.}~\bibnamefont
  {Blankenburg}},\ }\href {\doibase 10.1007/JHEP03(2011)133} {\bibfield
  {journal} {\bibinfo  {journal} {JHEP}\ }\textbf {\bibinfo {volume} {1103}},\
  \bibinfo {pages} {133} (\bibinfo {year} {2011})},\ \Eprint
  {http://arxiv.org/abs/1012.2697} {arXiv:1012.2697 [hep-ph]} \BibitemShut
  {NoStop}%
\end{thebibliography}%
\bibliographystyle{apsrev4-1}

\end{document}